\documentclass[aps, prb, amsmath,amssymb,amsfonts,twocolumn,superscriptaddress,floatfix,footinbib,altaffillsymbol]{revtex4-1}

\usepackage{amsmath}
\usepackage{amssymb}
\usepackage{dcolumn}
\usepackage[final]{graphicx}
\usepackage{upgreek}
\usepackage{color}
\usepackage[english]{babel}

\begin{document}
	
\title{Stochastic precession of the polarization in a polariton laser}
\author{V.~G.~Sala}\thanks{These authors equally contributed to the work.}
\affiliation{Laboratoire de Photonique et de Nanostructures (LPN), CNRS, Universit\'e Paris-Saclay, route de Nozay, F-91460 Marcoussis, France }
\affiliation{Laboratoire Kastler Brossel, Universit\'e Pierre et Marie Curie, \'Ecole Normale Sup\'erieure et CNRS, UPMC case 74, 4 place Jussieu, 75005 Paris, France}
\author{F.~Marsault}\thanks{These authors equally contributed to the work.}
\affiliation{Laboratoire de Photonique et de Nanostructures (LPN), CNRS, Universit\'e Paris-Saclay, route de Nozay, F-91460 Marcoussis, France }
\author{M.~Wouters}
\affiliation{Theory of Quantum and Complex Systems, Universiteit Antwerpen, Universiteitsplein 1, B-2610 Antwerpen, Belgium}
\author{E.~Galopin}
\affiliation{Laboratoire de Photonique et de Nanostructures (LPN), CNRS, Universit\'e Paris-Saclay, route de Nozay, F-91460 Marcoussis, France }
\author{I.~Sagnes}
\affiliation{Laboratoire de Photonique et de Nanostructures (LPN), CNRS, Universit\'e Paris-Saclay, route de Nozay, F-91460 Marcoussis, France }
\author{A.~Lema\^itre}
\affiliation{Laboratoire de Photonique et de Nanostructures (LPN), CNRS, Universit\'e Paris-Saclay, route de Nozay, F-91460 Marcoussis, France }
\author{J.~Bloch}
\affiliation{Laboratoire de Photonique et de Nanostructures (LPN), CNRS, Universit\'e Paris-Saclay, route de Nozay, F-91460 Marcoussis, France }
\affiliation{Physics Department, \'Ecole Polytechnique, F-91128 Palaiseau Cedex, France}
\author{A.~Amo}
\affiliation{Laboratoire de Photonique et de Nanostructures (LPN), CNRS, Universit\'e Paris-Saclay, route de Nozay, F-91460 Marcoussis, France }
\date{\today}
	
\begin{abstract}
Microcavity polaritons in the lasing regime undergo a spontaneous symmetry breaking transition resulting in coherent emission with a well defined polarization. The order parameter is thus a vector describing both the laser global phase and its polarization. Using an ultrafast single-shot detection technique we show that polariton lasing in GaAs-based microcavities presents a high degree of second order coherence ($g^{(2)}(\tau=0) \approx 1$) above threshold, and that the initial polarization is stochastic, taking any possible direction in the Poincar\'e sphere (linear, elliptical or circular). Once the polarization direction is established, subsequent oscillations of the emission probability witness the presence of an intrinsic polarization splitting. These results show the intricate polarization dynamics in the onset of polariton lasers.
\end{abstract}
	
	
\maketitle

\section{Introduction}

Polaritons in semiconductor microcavities have opened the door to the study of nonlinear phenomena in fluids of light~\cite{Carusotto2013}. One of their main properties is their ability to spontaneously accumulate in the same quantum state above a certain excitation density threshold, giving rise to the phenomenon of polariton lasing~\cite{Dang1998, Richard2005, Kasprzak2006, Balili2007, Christopoulos2007}. Similar to standard photon lasers~\cite{DeGiorgio1970, Graham1970}, the onset of polariton lasing is accompanied by the spontaneous breaking of the U(1) symmetry, resulting in emission with a high degree of temporal and spatial coherence~\cite{Kasprzak2006, Deng2007}. Additionally, polaritons are spinor quasiparticles with two possible projections of their internal spin along the growth axis of the microstructure, which map into right and left circularly polarized photons when leaking out of the cavity. Thus, in the presence of in-plane cylindrical symmetry, the order parameter of the polariton laser is a vector, and the spontaneous symmetry breaking results in the set up of a global and an internal phase, the latter defining the polarization of the emission.

This kind of spontaneous spin order emerges, for instance, in atomic Bose-Einstein condensates with several degenerate hyperfine levels and ferromagnetic-like interactions, resulting in the formation of spatially polarized domains~\cite{Sadler2006}. In the case of polaritons, interactions are in most situations antiferromagnetic (same spin interactions are repulsive and stronger than opposite spin ones)~\cite{Vladimirova2010,Paraiso2011, Takemura2014}, and it has been suggested that polariton lasing should then be linearly polarized, corresponding to the lowest energy (interacting) state~\cite{Shelykh2006}. This assumes the polariton laser being in thermodynamic equilibrium, which is not usually the case: the pump-dissipative dynamics might trigger lasing in excited states~\cite{Wouters2008b}. Additionally, polariton-polariton interactions are rather weak at threshold (the interaction energy is much smaller than the linewidth). In line with these two arguments, out-of-equilibrium and weak interactions, Read and coworkers predicted that close to threshold, the initial polarization should be completely random, taking any possible value in the Poincar\'e sphere with equal probability, including circular and elliptical polarizations~\cite{Read2009a}. The same phenomena is expected in a photon laser in semiconductor VCSELs.

Despite the key role of the polarization initialization and subsequent evolution in the symmetry breaking physics in microcavities, experiments have not yet addressed the polarization dynamics in the onset of polariton lasing. The reason is the required high temporal resolution, on the order of the polariton coherence time --being as low as a few picoseconds below threshold--, in combination with single shot experiments capable of resolving the initial polarization on each experimental realization. Experiments under continuous wave excitation have shown polariton lasing whose polarization was pinned to the crystallographic axis or to local spatial inhomogeneities~\cite{Klopotowski2006, Kasprzak2006, Krizhanovskii2006, Kasprzak2007}. This situation results in a classical bifurcation to circularly polarized states under strong pumping, when polariton-polariton interactions are relevant~\cite{Ohadi2015}. In the pulsed regime, experiments have been analyzed by integrating the emission over its whole duration in Refs.~\onlinecite{Baumberg2008, Ohadi2012}. Those works showed evidence of the stochastic initial polarization direction in a polariton laser, but the temporal dynamics was not accessed. The vector symmetry breaking physics of a polariton laser still lacks ultrafast experimental reports.

In this article we use a single-shot ultrafast detection technique based on a streak camera with a time resolution of $4 \,\mathrm{ps}$ to measure the polarization dynamics of a GaAs/AlGaAs polariton laser via the statistics of the emitted intensity~\cite{Ueda1988, Wiersig2009}. The second order coherence function of the total emitted photons at zero delay ($g^{(2)}(\tau=0)$) rapidly decreases to 1 above the condensation threshold, showing that the statistics of the polariton laser emission is poissonian. When the emission is selected in polarization, we observe that the initial polarization is stochastic, taking any possible direction in the Poincar\'e sphere (linear, elliptical or circular polarization). Subsequent oscillations of the second order correlation witness the rotation of the polarization around the direction of an intrinsic linear polarization splitting present in our samples. When the initial polarization is circular, polariton interactions counteract the polarization splitting and preserve the initial polarization. Thus, contrary to the extended idea that the initial polarization of the polariton laser should be linear~\cite{Shelykh2006}, we show that it can take any value, in agreement with the prediction of Read et al.~\cite{Read2009a} in the weakly interacting regime.

The paper is organized as follows. In Sec.~\ref{SecExp} we describe the micropillar and planar microcavity samples employed in our studies, as well as the experimental set-up; Section~\ref{SecCoh} shows the degree of second order coherence measured for the total emitted intensity in both samples; Section~\ref{SecInitPol} addresses the initial polarization distribution of the polariton laser; Section~\ref{SecPolDyn} is devoted to the observaton of the polarization precession along with the model that describes the experimental findings.

\begin{figure}[t]
	{\includegraphics[width=\columnwidth]{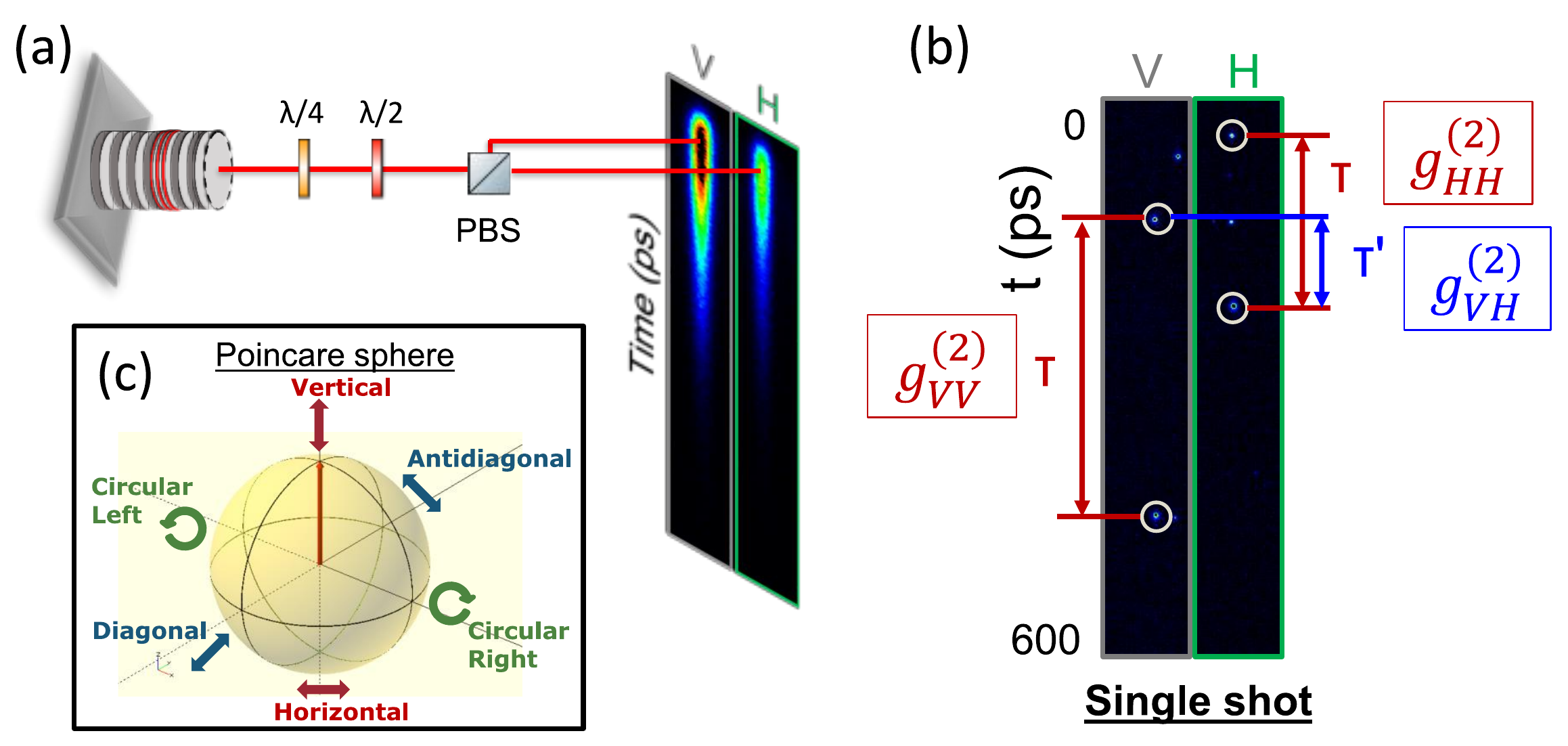}}
	\caption{
		(a) Scheme of the experimental setup, streak camera images integrated over fifty million excitation pulses.
		(b) Emission measured as a function of time in the single-shot mode of the streak camera and the six Stokes parameters.
		(c) Scheme of the Poincare sphere.
	}
	\label{FigS1}
\end{figure}

\section{Experimental system}
\label{SecExp}

Our sample is grown by molecular beam epitaxy and consists of a $ \lambda / 2 \ \mathrm{Ga}_{0.05}\mathrm{Al}_{0.95}\mathrm{As} $ cavity surrounded by two $\mathrm{Ga}_{0.8}\mathrm{Al}_{0.2}\mathrm{As} / \mathrm{Ga}_{0.05}\mathrm{Al}_{0.95}\mathrm{As} $ Bragg mirrors with $28$ (top) and $40$ (bottom) pairs. The nominal quality factor of the cavity is $Q=72000$. Twelve $7 \,\mathrm{nm}$ GaAs quantum wells are positioned on the three central anti-nodes of the electromagnetic field, resulting in a Rabi splitting of $15 \,\mathrm{meV}$. Experiments are realized both in the as-grown planar microcavity and in a pillar of $3 \,\mu \mathrm{m}$ diameter fabricated using electron beam lithography and dry etching. In both samples the detuning between the cavity mode and exciton energy is $+3 \,\mathrm{meV}$.

Photoluminescence experiments are performed at 5K using a pulsed Ti:sapphire laser delivering 3~ps pulses at a repetition rate of 82~MHz. The laser energy is tuned $100~\,\mathrm{meV}$ above the polariton resonance. A microscope objective (NA=0.65) is used both to focus the laser on a $2 \,\mu \mathrm{m}$ spot and to collect the emission, which is time resolved using a streak camera operated in a single-shot mode. For photoluminescence measurements, the emitted signal is dispersed in a monocromator before reaching the streak camera, resulting in a time resolution of 8~ps. In intensity correlation measurements we use instead a broadband longpass filter with a cutoff wavelength of 750~nm, which prevents the excitation laser at 735~nm from reaching the detector. In this case the temporal resolution is improved to $4 \,\mathrm{ps}$. In both configurations the emission is analyzed along the six Stokes polarization axis [Fig.~\ref{FigS1}(c)] with the use of a $\lambda/4$ and $\lambda/2$ waveplates in combination with a polarizing beamsplitter: The emitted photons are separated into two beams of arbitrary orthogonal polarizations, which are simultaneously imaged onto two different positions of the entrance slit of the streak camera [Fig.~\ref{FigS1}(a)]. The total emitted intensity as a function of time is retrieved by adding the signal from two orthogonal polarizations.

\begin{figure}[t]
	{\includegraphics[width=\columnwidth]{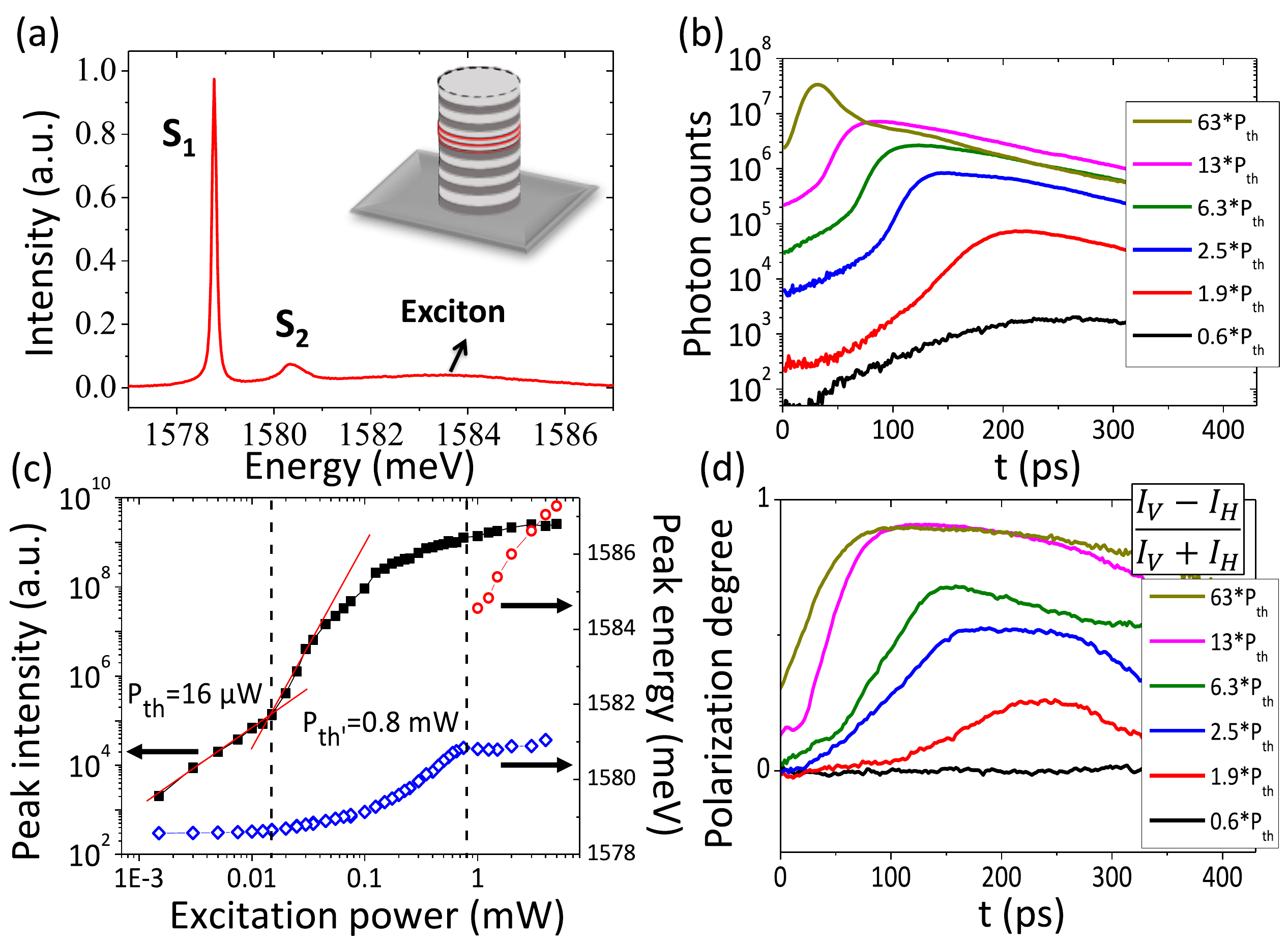}}
	\caption{
		(a) Photoluminescence spectrum of the micropillar at $P=0.1 \,P_{th}$.
		(b) Polariton emission ($S_1$) as a function of time for increasing excitation power.
		(c) Peak intensity (black squares) and peak energy of the polariton emission (blue diamonds) and of the photon emission (red circles) as a function of the excitation power. The vertical dashed lines stand for the two thresholds $P_{th}=16 \, \mu \mathrm{W}$ and $P_{th\textsc{\char13}}=0.8 \,\mathrm{mW}$.		
		(d) Averaged Vertical-Horizontal degree of linear polarization as a function of time for the excitation powers shown in (b).
	}
	\label{Fig1}
\end{figure}

Let us first characterize the lasing regimes and their dynamics in the micropillar. The time integrated spectrum measured at low power is shown in Fig.~\ref{Fig1}(a). It displays two polariton modes. The measured linewidth of the lowest energy one, $\mathrm{S}_1$, is $100 \,\mu \mathrm{eV}$, larger than that expected from the $Q$ factor ($\sim 22~\mu \mathrm{eV}$). This broadening is attributed to spectral wandering induced by fluctuations in the charge environment of the quantum wells~\cite{Besombes2002}.

The dynamics of the polariton emission for increasing excitation power is depicted in Fig.~\ref{Fig1}(b). Each trace shows the accumulation of about fifty million realizations in the streak camera. Above the threshold power $P_{th}=16 \,\mu \mathrm{W}$, the emission is fully dominated by $\mathrm{S}_1$, a sharp increase of the intensity is observed [Fig.~\ref{Fig1}(c)], and the dynamics accelerates [Fig.~\ref{Fig1}(b)]. This behavior is the signature of stimulated relaxation of polaritons into $\mathrm{S}_1$ and the onset of polariton lasing. A second threshold appears at $P_{th\textsc{\char13}}=0.8 \,\mathrm{mW} = 50 \,P_{th}$, characterized by a fast emission at short time delays occurring at higher energy. It corresponds to conventional photon lasing as the system reaches the weak coupling regime~\cite{Houdre1995, Jahnke1997, Butte2002a, Bajoni2008}.

The polarization degree of the emission in the Vertical/Horizontal axis $ \rho_{VH} = \frac{I_V-I_H}{I_V+I_H} $ is reported as a function of time in Fig.~\ref{Fig1}(d), where $I_V$ ($I_H$) is the emission intensity in the Vertical (Horizontal) polarization, corresponding to the crystalline axis of the sample. Below threshold, the degree of polarization is negligible. Above $P_{th}$, the polariton lasing emission is vertically polarized up to $\rho_{VH}=0.9$, while the diagonal and circular polarization degrees are close to zero for the whole power range. In these experiments, performed by the accumulation of several million realizations, the polarization of the emission appears to be pinned along the vertical direction, indicating the presence of a polarization splitting induced by strain along the crystalline axis of the sample~\cite{Kasprzak2006}. Since the emission below threshold is unpolarized, the polarization splitting must be smaller than the apparent linewidth of the polariton mode in the linear regime [Fig.~\ref{Fig1}(a)]. Similar intensity and polarization dynamics are observed in the as-grown planar sample (not shown).

\section{Degree of second order coherence}
\label{SecCoh}

An important characteristic of lasing emission is the second order correlation at zero delay $g^{(2)}(\tau=0)$: for a conventional single-mode laser, we expect a monotonous transition from a value of 2 (thermal emission) to 1 (coherent-poissonian emission) when crossing the lasing threshold. This transition has been experimentally studied in microcavity lasers in the weak~\cite{Assmann2009, Assman2010} and strong coupling regimes~\cite{Adiyatullin2015, Kim2016}. In our experiment, below threshold, the thermal emission comes from two independent polarized modes, and the expected value is 1.5 instead of 2 if no polarization selection is performed in the detector. 
For photon-correlation experiments we use a streak camera with a single-shot resolution of $4 \,\mathrm{ps}$ following the method of Wiersig et al.~\cite{Wiersig2009}. In this technique, the time of arrival of each photon is measured in order to build the second order correlation function:

\begin{align}
g^{(2)}_{total}(t,\tau) = \frac{ \langle \hat{a}^{\dagger}(t) \hat{a}^{\dagger}(t+\tau) \hat{a}(t+\tau) \hat{a}(t) \rangle }{\langle \hat{a}^{\dagger}(t) \hat{a}(t)\rangle \langle \hat{a}^{\dagger}(t+\tau) \hat{a}(t+\tau) \rangle},
\end{align}

\noindent where $\hat{a}^{\dagger}(t)$ is the creation operator of photons emitted by the micropillar at time $t$, and the brackets indicate statistical averages. The subscript "total" indicates that we consider photons regardless of their polarization. This function accounts for the probability of conditional emission of a photon at time $t+\tau$ given the emission of a photon at time $t$.

\begin{figure}[t]
	{\includegraphics[width=\columnwidth]{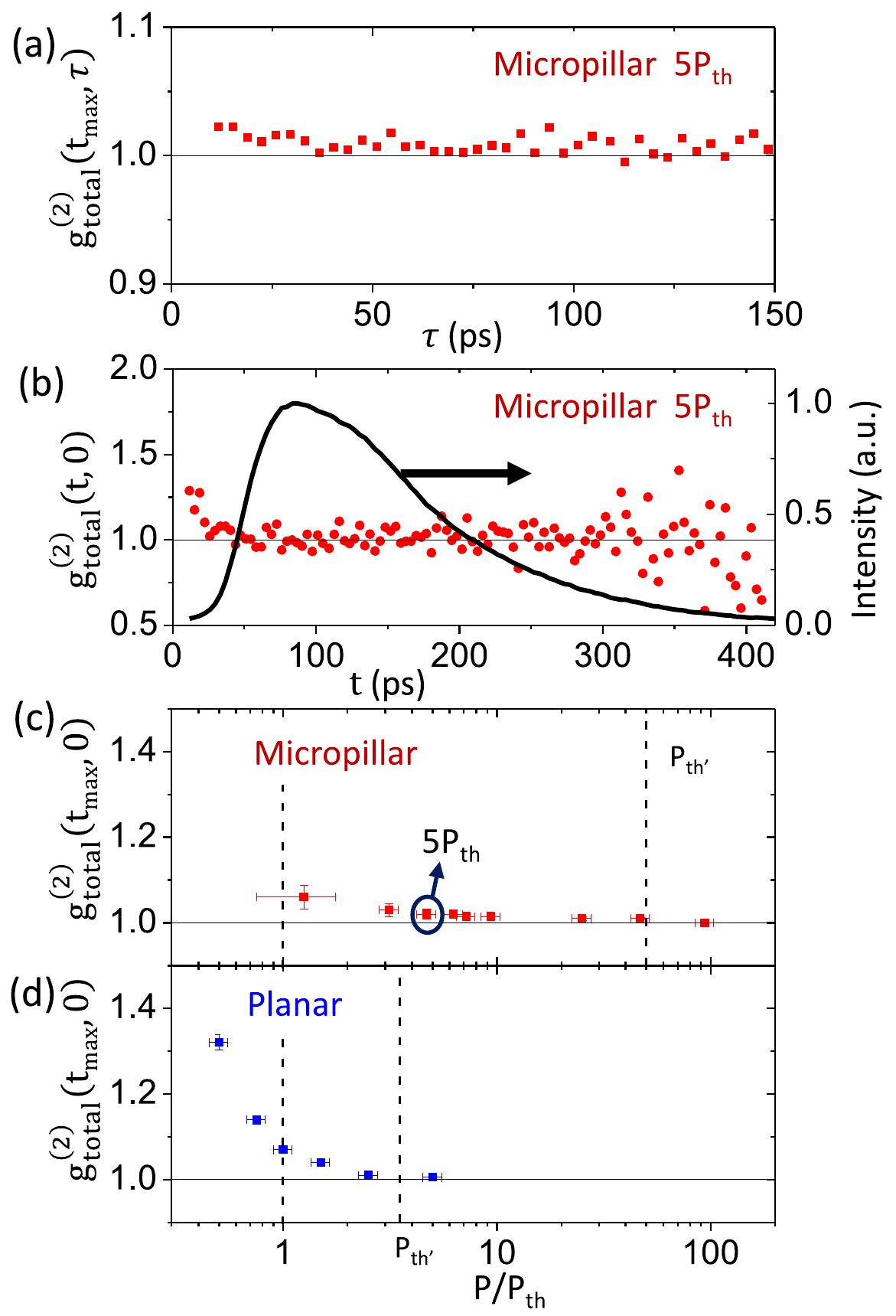}}
	\caption{
		(a) Second order correlation function of the total polariton emission at $t_{max}$ as a function of the delay $\tau$ ($g^{(2)}_{total}(t_{max},\tau)$) measured at $P=5~P_{th}$ in the micropillar. (b) Zero-delay autocorrelation function $g^{(2)}_{total}(t,0)$ (dots) as a function of time after arrival of the excitation pulse at $P=5~P_{th}$. The solid line shows the emitted intensity as a function of time. (c)-(d) Zero-delay autocorrelation function $g^{(2)}_{total}(t,0)$ at $t_{max}$ as a function of excitation density for the micropillar (a) and the planar microcavity (b).
	}
	\label{Fig2}
\end{figure}

Figure~\ref{Fig2}(a) shows $g^{(2)}_{total}(t_{max},\tau)$ at $P=5 P_{th}$, that is, the correlation function when the first photon arrives at the time of the maximum of the emission $t_{max}$ and the second photon at a later time $t_{max}+\tau$. Spectrally resolved measurements, show that at this power the emission is fully dominated by the polariton state $S_1$. While the time resolution of the streak camera technique is 4~ps, the shortest delay between photons that we can measure is 10~ps. This is related to the effective size of the photons in the streak camera detector and the fact that the arrival of two photons in the same pixel gives the same signal as the arrival of one single photon (see Supporting Online Material of Ref.~\onlinecite{Assmann2009}). In the rest of the paper we will refer to this resolution limited value of $g^{(2)}$ at the shortest delay $\tau$ as $g^{(2)}_{total}(t,0)$. Figure~\ref{Fig2}(a) shows a value of $g^{(2)}_{total}(t_{max}, 0)$ of 1.02 and a subsequent decrease towards 1.00 at longer delays $\tau$, with a decay time of 40~ps. 

By selecting the value of $g^{(2)}_{total}$ at the shortest time delay $\tau$, our technique allows us tracking the value of $g^{(2)}_{total}(t,0)$ as a function of time $t$ after the arrival of the excitation pulse. This is shown in Fig.~\ref{Fig2}(b) for the emission at $P=5 P_{th}$. As soon as the polariton laser switchs on, at around $t=25$~ps, $g^{(2)}_{total}(t,0)$ decreases from around 1.3 to 1.0, as expected from a coherent source and it stays close to one not only at $t_{max}$ but all along its emission. At long times, when the emitted intensity gets low and the lasing effect ceases, we would expect an increase of $g^{(2)}_{total}(t,0)$ associated to the loss of coherence~\cite{Assman2010}. However the low photon yield prevents us from studying this situation.

The dependence in excitation power of the equal-time correlations at $t_{max}$ is summarized in Fig.~\ref{Fig2}(c). To reduce the error bar, we plot the measured value of $g^{(2)}_{total}(t,0)$ averaged over emission times between $t = t_{max}- 20~\mathrm{ps}$ and $t = t_{max}+ 20~\mathrm{ps}$. As the first threshold $P_{th}$ is crossed, $g^{(2)}_{total}(t_{max},0)$ decreases from a value lower than 1.5 towards 1.0, and it remains close to 1.0 above $P_{th}$. The same behavior is observed for the planar cavity [Fig.~\ref{Fig2}(c), using an excitation spot of $15 \,\mu \mathrm{m}$ in diameter]. At $2.5 P_{th}$, $g^{(2)}_{total}(t_{max},0)$ amounts to $1.01$, similar to the value reported for a zero-dimensional monomode polariton cavity~\cite{Kim2016}.

Contrary to previous reports in planar cavity structures with a lower quality factor~\cite{Horikiri2010, Tempel2012}, we do not observe any increase of the noise with increasing excitation density. $g^{(2)}_{total}(t_{max},0)$ remains $\simeq 1$ above $P_{th}$, even when increasing the excitation density above the threshold for photon lasing in the weak coupling regime ($P_{th\textsc{\char13}}$).

These observations are in agreement with quantum Monte-Carlo based calculations including weak polariton-polariton interactions~\cite{Wouters2009a}, and recent experimental reports~\cite{Adiyatullin2015}, and show the negligible role of interactions in the intensity correlations of a polariton laser.

\section{Initial polarization}
\label{SecInitPol}

The single shot experimental set-up allows us studying the initial stochastic polarization of the polariton laser and its subsequent dynamics.

To do so, we separate the emission in two beams of opposite polarization that are imaged at two different positions of the streak camera. As shown in Fig.~\ref{FigS1}(b), this configuration allows the study of the conditional detection of a photon in polarization $Y$ at $t+\tau$ given the emission of a previous photon in the opposite polarization $X$ at time $t$. This is casted in the following cross-correlation function:

\begin{align}
g^{(2)}_{XY}(t,\tau) = \frac{ \langle \hat{a}^{\dagger}_{X}(t) \hat{a}^{\dagger}_{Y}(t+\tau) \hat{a}_{Y}(t+\tau) \hat{a}_{X}(t) \rangle }{\langle \hat{a}^{\dagger}_{X}(t) \hat{a}_{X}(t)\rangle \langle \hat{a}^{\dagger}_{Y}(t+\tau) \hat{a}_{Y}(t+\tau) \rangle},
\end{align}

\noindent where {X} and {Y} stands for either Horizontal (H) and Vertical (V), or Diagonal (D) and Antidiagonal (A), or Circular Left (L) and Circular Right (R) polarization of detection.

The auto-correlation function at zero delay, $g^{(2)}_{{XX}}(t,0)$, is determined by the probability distribution for the system to start lasing in a given polarization~${X}$. In the case of an initial random distribution, a value of $g^{(2)}_{{XX}}(t,0)=1.33$ is expected for every polarization. If, differently, the initial polarization is restricted to be linear as suggested in Ref.~\onlinecite{Shelykh2006}, $g^{(2)}_{{XX}}(t,0)=1.5$ for any linear polarization direction and $g^{(2)}_{{XX}}(t,0)=1$ for circular polarization (see Appendix~\ref{AppI}).

\begin{figure}[t]
	{\includegraphics[width=0.9\columnwidth]{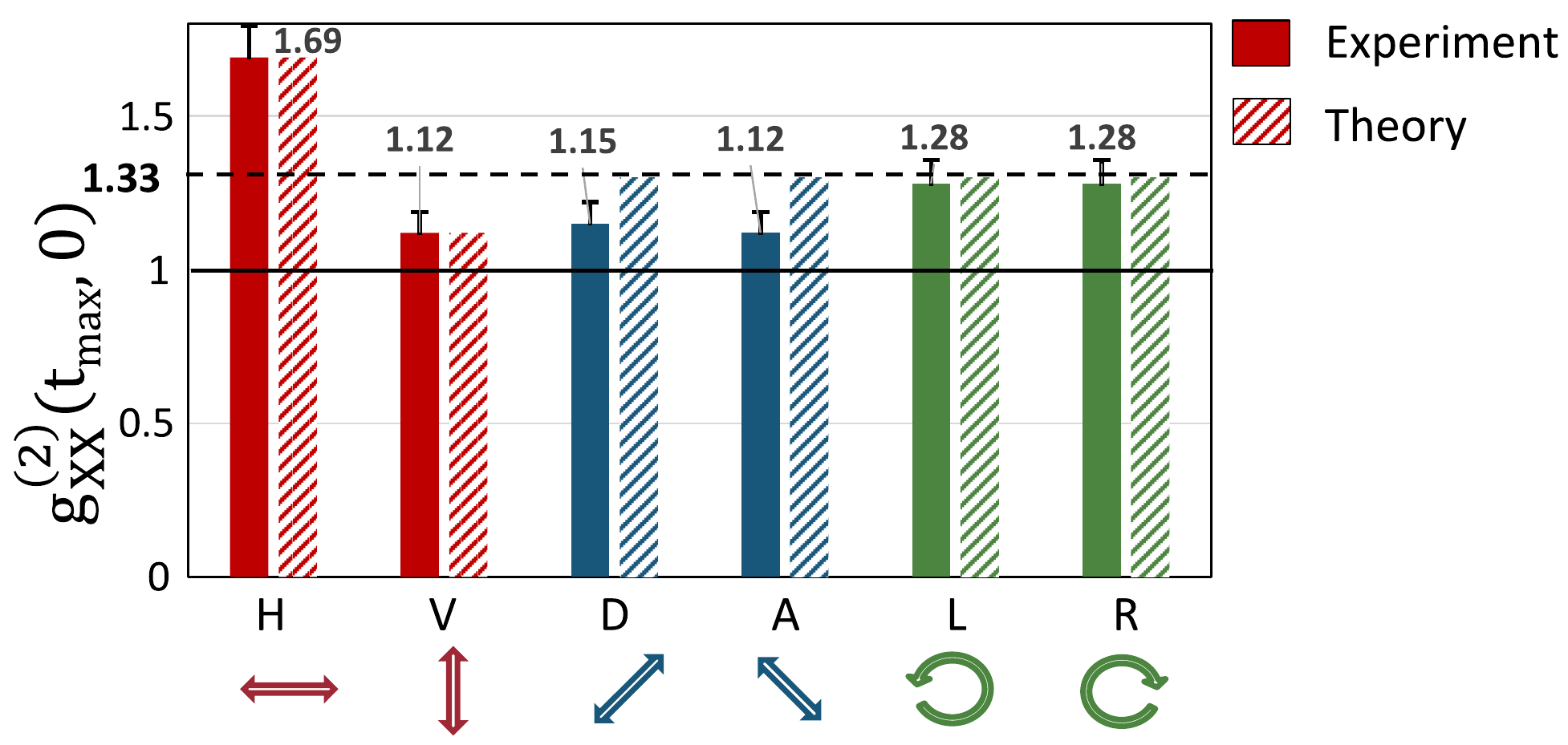}}
	\caption{
		(a) Experimental (full) and theoretical (stripped) values of the auto-correlation function at $\tau=0$ in the Horizontal-Vertical (red), Diagonal-Antidiagonal (blue), and Circular (green) polarizations of the pillar emission at $P=5~P_{th}$. Error bars are shown on top of each column.
	}
	\label{Fig3}
\end{figure}

The measured value of $g^{(2)}_{{XX}}(t_{max},0)$ for the micropillar at $P=5P_{th}$ is shown in Fig. \ref{Fig3} (full bars) for each polarization axis. 250~000 emission pulses have been recorded. Firstly, 
the fact that $g^{(2)}_{LL/RR}(t_{max},0)$ in the circular axis is larger than 1 shows that the polarization distribution is not exclusively linear, the system having a non-zero probability of starting lasing with circular polarization. 

\begin{figure*}[t!]
	{\includegraphics[width=0.6\textwidth]{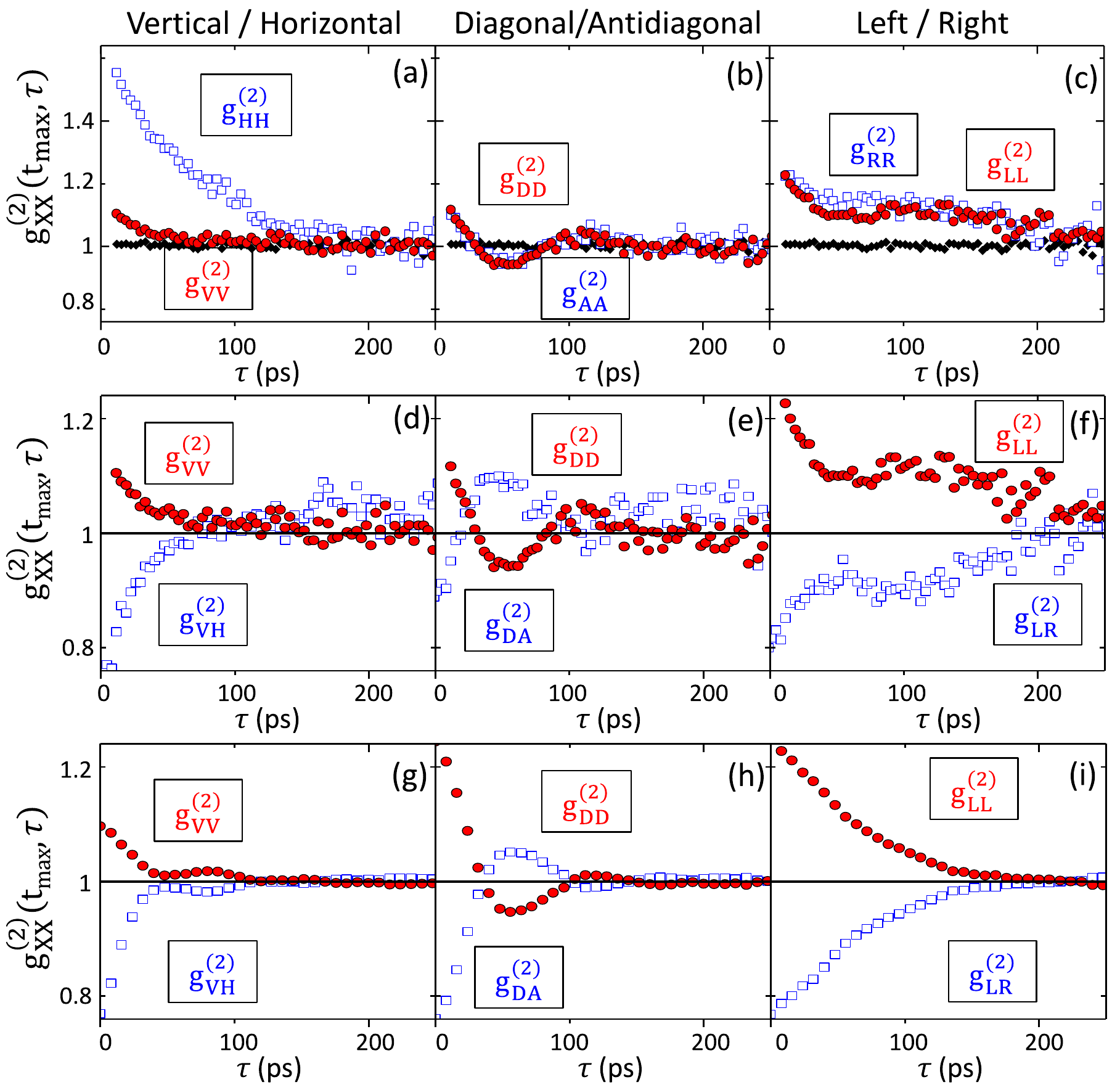}}
	\caption{
		(a)-(c) Polarization resolved auto-correlation $g^{(2)}_{{XX}}(t_{max},\tau)$ measured for the micropillar at $P=5~P_{th}$. The black diamonds show the auto-correlation of all the emitted photons (without any selection in polarization). (d)-(f) Auto-correlation $g^{(2)}_{{XX}}(t_{max},\tau)$ (red full circles, same data as in (a)-(c) for V, D and L polarizations) and cross-correlation $g^{(2)}_{{XY}}(t_{max},\tau)$ (blue open squares). 
		(g)-(i) Corresponding Monte Carlo simulations.
	}
	\label{Fig4}
\end{figure*}

Second, the strong disparity between horizontal and vertical polarization confirms the presence of a polarization splitting along the crystallographic axis of the system. The values of $g^{(2)}_{{XX}}(t_{max},0)$ are qualitatively reproduced in Fig.~\ref{Fig3} (striped bars) through a simple model where the following normalized probability distribution is assumed: V polarization is 4.5~times more likely than H polarization, and the likelihoods of D, A, L or R are the same (see Appendix~\ref{AppI} for the calculation linking these probabilities to the simulated initial $g^{(2)}_{{XX}}(t_{max},0)$). Note that the lower probability to lase in the H direction results in an increase of the measured $g^{(2)}_{{HH}}(t_{max},0)$ with respect to $g^{(2)}_{{VV}}(t_{max},0)$. The lower values of D and A in the experiment with respect to the model suggest that their respective likelihoods may be slightly different than for L and R polarizations.

These results show that despite the presence of an intrinsic polarization splitting that favours lasing along one of the polarization eigenstates (V in our case), the initial polarization of the emission presents a stochastic character, and it can occur in any polarization state. This means that the dynamics for the onset of lasing are faster than or on the order of $\hbar / \Delta_{VH}$, where $\Delta_{VH}$ is the intrinsic polarization splitting. Analogous results are obtained in the planar microcavity (see the short delay times $\tau$ in Fig.~\ref{FigS4}).

\section{Polarization dynamics}
\label{SecPolDyn}

The polarization dynamics after each initialization of the polariton laser can be studied by tracking $g^{(2)}_{{XX}}(t_{max},\tau)$ as a function of~$\tau$. 
Figure~\ref{Fig4}(a-c) shows the measured auto-correlation of the photons emitted by the micropillar at $P=5P_{th}$ for different polarizations, as well as without any selection in polarization (black diamonds). Figure~\ref{Fig4}(d-f) depicts the auto-correlation for Vertical, Diagonal and Left circular polarizations [red dots show the same data as in (a)-(c)] and the cross-correlation function $g^{(2)}_{{XY}}(t_{max},\tau)$ between opposite $XY$ polarizations (blue squares) following the procedure described above and depicted in Fig.~\ref{FigS1}(b).

Let us first consider the Vertical polarization direction. While the auto-correlation is constant and equal to 1 for the total emission (black dots) [Fig. \ref{Fig4}(a)], $g^{(2)}_{VV}(t_{max},\tau)$ [resp. $g^{(2)}_{VH}(t_{max},\tau)$] shows a monotonous decay [resp. increase] from 1.12 [resp. 0.70] towards 1.00 [Fig. \ref{Fig4}(d)]. This is a consequence of this axis being parallel to the polarization of an eigenstate of the system: if the laser starts with a polarization along this axis, it preserves it for the whole duration of the emission. 
The decay of $g^{(2)}_{VV}(t_{max},\tau)$ from its initial value towards 1 reflects the spin decoherence induced by interactions with reservoir excitons~\cite{Ohadi2015}.

If the initial polarization is diagonal [Fig. \ref{Fig4}(e)], oscillations of $g^{(2)}_{{DD}}(t_{max},\tau)$ and $g^{(2)}_{{DA}}(t_{max},\tau)$ above and below 1 are observed. An initial diagonal polarization can be seen as the coherent superposition of the two split polarization eigenstates (Vertical-Horizontal). As time evolves, the frequency difference between the two states results in a continuously running phase difference, evidenced by the precession of the polarization around the VH axis in the Poincar\'e sphere from diagonal to circular, antidiagonal, circular, diagonal,.... The precession results in oscillations of the probability of measuring a second photon parallel to the Diagonal axis [Fig.~\ref{Fig4}(e), red full circles], which is anti-correlated with those observed when the first photon is diagonal and the second is anti-diagonal [Fig.~\ref{Fig4}(e), blue open squares]. 

An analogous precession should also occur for the circular polarization. However, Fig.~\ref{Fig4}(c), (f) shows that while $g^{(2)}_{{LL}}(t_{max},\tau)$ presents some oscillations, the correlation function stays above 1. This means that if lasing starts in the left circular polarization, it stays globally left circularly polarized (there are no oscillations between L and R polarizations). This can be understood accounting for the spin anisotropy of polariton-polariton interactions, which are much stronger for same spin than for opposite spin polaritons~\cite{Vladimirova2010, Takemura2014}. Thus, if the polariton laser is initially circularly polarized, the intrinsic linear polarization splitting is partially screened and the polarization precesses around new nonlinear eigenstates of elliptical polarization~\cite{Read2009a}, which are determined by the spontaneous initial population imbalance between the two circular polarizations. This effect is known as self-induced Larmor precession. 
The observed behaviour indicates that polariton interactions, instead of favoring linearly polarized lasing, actually help preserving the degree of circular polarization.

The dynamics of $g^{(2)}_{{XX}}(t_{max},\tau)$ can be well reproduced assuming the following two coupled equations of motion of the polariton field in the circular polarization basis:
\begin{align}
i \frac{d}{dt} \Psi_{L(R)} = \alpha_{1} |\Psi_{L(R)}|^2 \Psi_{L(R)} +(-) \frac{i}{2} \Delta_{VH} \Psi_{R(L)}+\sqrt{\sigma} \xi(t),
\end{align}

\noindent where $\alpha_{1}$ is the same-spin polariton-polariton interaction constant (we neglect opposite spin interactions), $\Delta_{VH}$ is the intrinsic polarization splitting along the Vertical-Horizontal axis. The diffusion term $\sqrt{\sigma} \xi(t)$ accounts for the randomization of the polarization due to fluctuations, and it results in a decay of the envelope of $g^{(2)}_{{XX}}(t_{max},\tau)$ from its initial value towards 1. Simulations of the evolution of $g^{(2)}_{{XX}}(t,\tau)$ for the modelled initial stochastic polarization distributions shown in Fig.~\ref{Fig3}, with $\alpha_{1} (|\Psi_{L}|^2+|\Psi_{R}|^2)= 80 \,\mu \mathrm{eV}$, $\Delta_{VH}=16 \,\mu \mathrm{eV}$, and a spin diffusion coefficient $\sigma /|\Psi|^2=0.0025 \,\mathrm{ps}^{-1}$ reproduce quantitatively the observed oscillations, as shown in Fig.~\ref{Fig4}(g)-(i). These fitting parameters agree well with the interaction energy estimated from the experiment: Assuming a value of $\alpha_{1} = 2~\mu$eV $\mu$m$^{2}$,\cite{Ferrier2011} from the total emitted intensity we estimate an interaction energy of 85~$\mu$eV at $t_{max}$ and $P=5~P_{th}$ (see Appendix~\ref{AppII} for the estimate procedure). 
As both the interaction energy and the intrinsic polarization splitting are of the same order of magnitude, the oscillations reflect the interplay between the self-induced Larmor precession and intrinsic splitting-induced oscillations.

\begin{figure}[t]
	{\includegraphics[width=\columnwidth]{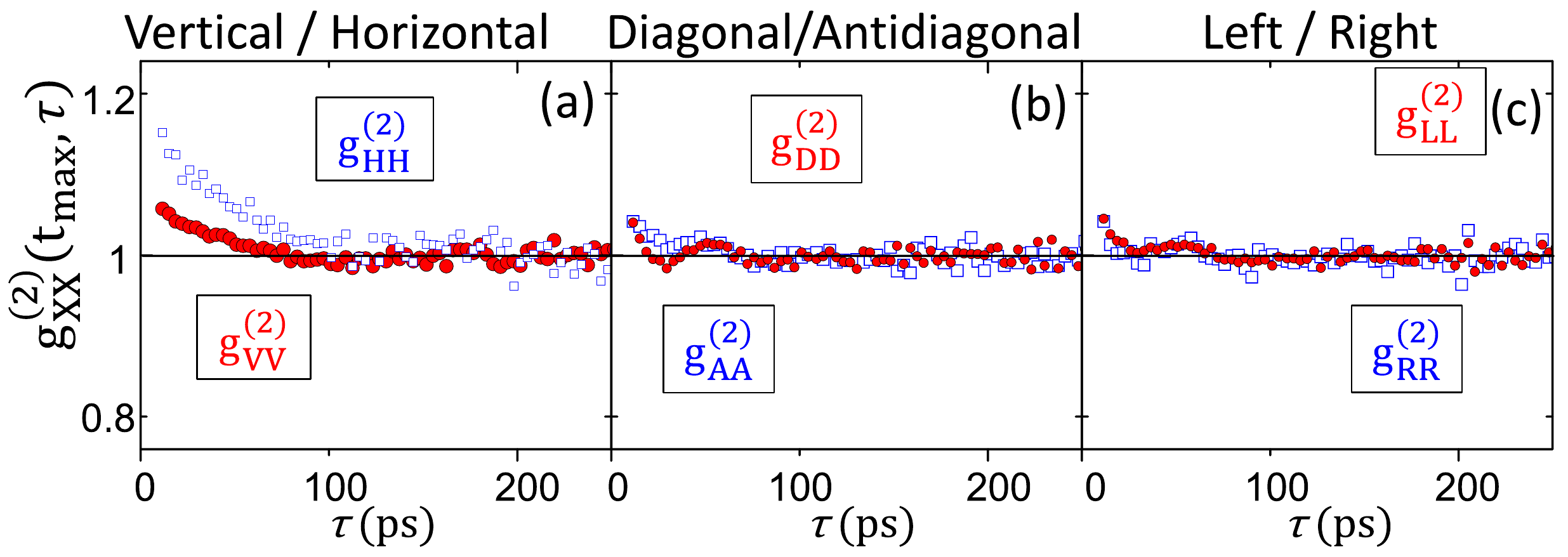}}
	\caption{
		(a)-(c) Polarization selected auto-correlation $g^{(2)}_{XX}(t_{max},\tau)$ for V, D and L (red circles) and for H, A and R (blue squares) for the planar microcavity emission at $P=1.5P_{th}$.
	}
	\label{FigS4}
\end{figure}

In the same direction it is interesting to analyze the polarization dynamics in the case of the planar microcavity. Figure \ref{FigS4}(a-c) shows the measured auto-correlation function in the planar cavity for the six considered polarizations at $P=1.5P_{th}$, in the polariton lasing regime. Similarly to the behavior found in the pillar cavity, $g^{(2)}_{VV}(t_{max},\tau)$ and $g^{(2)}_{HH}(t_{max},\tau)$ [Fig.~\ref{FigS4}(a)] show a monotonous decay from 1.07 and 1.19, respectively, towards 1.00. The different value at zero delay of $g^{(2)}_{XX}(t_{max},0)$ for Vertical and Horizontal directions evidence again the existence of an intrinsic polarization splitting along the crytallographic axis. If the polariton laser starts with any other polarization, we observe oscillations of $g^{(2)}_{XX}(t_{max},\tau)$, as shown for D, A, L and R in Fig. \ref{FigS4}(b)-(c)].

The amplitude of the oscillations around $g^{(2)}_{XX}=1$ is of the order of 0.05, smaller than in the pillar cavity ($\sim 0.15$). This difference may arise from the coexistence of several spatial polariton modes at threshold in the planar cavity, while in the micropillar the photonic confinement results in just two orthogonally polarized modes with the same spatial profile (the $\mathrm{S}_1$ modes). From the quality factor of the cavity we estimate the number of polariton modes in the emission spot size to be on the order of 7 (see Appendix~\ref{AppII}). At the onset of lasing, different points within the spot might start lasing both with a different spontaneous phase~\cite{Wouters2009a} and a different polarization in the equivalent of the Kibbel-Zurek mechanism for a vectorial order parameter. While the phase and the polarization are expected to get homogeneous over the whole spot some time after the initialization of the laser, the initial polarization distribution in different modes might result in the reduced value of $g^{(2)}_{XX}(t_{max},\tau)$ observed in the planar microcavity.

A second difference between the planar cavity and the micropillar observations is the fact that in the planar cavity, $g^{(2)}_{RR}(t_{max},\tau)$ and $g^{(2)}_{LL}(t_{max},\tau)$ show oscillations that cross zero [Fig.~\ref{FigS4}(c)], while they stay above zero in the micropillar [Fig.~\ref{Fig4}(c)]. This indicates that interactions play a negligible role in the planar cavity case, and that the self-induced Larmor precession mechanism does not participate in the dynamics. We can check this hypothesis by estimating the interaction energy in a similar way as for the micropillar. For the planar cavity at $P=1.5P_{th}$ we estimate a total polariton interaction energy of $0.54 \,\mu \mathrm{eV}$ (see Appendix~\ref{AppII}). This value can be compared to the polarization splitting that results in the oscillation observed in Figs.~\ref{FigS4}(b)-(c). From the oscillation period of 60~ps, we deduce an energy splitting of $34 \,\mu \mathrm{eV}$, much larger than the estimated interaction energy. We can thus identify the oscillation period with the splitting $\Delta_{VH planar}$: In the case of the planar cavity the precession of the polarization depends only on the intrinsic polarization splitting along the Vertical/Horizontal axis. In this situation, the Diagonal/Antidiagonal and Left/Right circular polarization axis are equivalent, and a similar oscillation behavior is expected as observed in Figs.~\ref{FigS4}(b)-(c).

\section{Summary}

Our results show that polariton lasers present a degree of second order coherence very close to 1, as expected from a standard laser source. The initial polarization is not set by the intrinsic polarization splitting (for the moderate values present in our samples). On the contrary, it results from spontaneous symmetry breaking inherent to the lasing process, and it can give rise to lasing in any polarization state. The intrinsic splitting has two effects: (i) it favors lasing polarized in the direction parallel to one of the splitting axis and, (ii) it gives rise to the precession of the polarization after lasing. This behavior might explain the low value of the total degree of polarization ($< 0.3$) and the negligible circular polarization reported by Ohadi et al.~\cite{Ohadi2012} in pulsed single shot measurements integrated in time. The dynamics evidenced in our experiments reflects the universal behavior of symmetry breaking in microcavity lasers. The stochastic initialization of the polarization and the subsequent precession are not exclusive of the polariton system and should also be present in standard photon lasers based on VCSELs~\cite{Gerhardt2011}.

This work was supported by the Agence Nationale de la Recherche project Quandyde (Grant No. ANR-11-BS10-001), the LABEX Nanosaclay project Qeage (Grant No. ANR-11-IDEX-0003-02), the FP7 ITN "Clermont4" (235114), the French RENATECH network, the ERC grant Honeypol and the EU-FET Proactiv grant AQUS (Project No. 640800).

\appendix

\section{Initial polarization}
\label{AppI}

In order to model the measured values of the zero time delay auto-correlation functions $g^{(2)}_{XX}(t_{max},0)$ along the different polarization axis (Fig.~\ref{Fig3}), we employ a statistical description of the initial polarization on the Poincare sphere [Fig. \ref{FigS1}(c)]. The eigenstates of the system are given by the polarization splitting along the vertical ($|V \rangle$) and horizontal ($|H \rangle$) axis. A general polariton lasing state can be written:
\begin{align}
|\psi \rangle = A \left[\cos\left(\frac{\theta (t)}{2}\right) |V \rangle + \exp^{i\phi (t)} \sin\left(\frac{\theta (t)}{2}\right) |H \rangle \right]
\end{align}
where $\theta (t)$ and $\phi (t)$ are, respectively, the polar and azimuthal angles in the Poincar\'e sphere, and $A$ is a normalization constant. $\phi = 0$ corresponds to linear polarization along an axis given by $\theta$. When the polarization is selected in our measurement, the wave function is projected onto the 'detector state': 
$|Det \rangle =\cos(\frac{\alpha}{2}) |V \rangle + \exp^{i\beta} \sin(\frac{\alpha}{2}) |H \rangle $, parametrized by $\alpha$ and $\beta$, which account for the positions of the $\lambda/2$ and $\lambda/4$ waveplates.

The density measured by the detector is then: 
$n(\alpha, \beta, \theta, \phi) = |\langle Det|\psi \rangle|^2 $,
and the measured auto-correlation function is given by: 

\begin{align}
g^{(2)}(t,\tau) = \frac{\langle n[\alpha, \beta, \theta(t+\tau), \phi(t+\tau)] n[\alpha, \beta, \theta(t), \phi(t)] \rangle}{\langle n[\alpha, \beta, \theta(t+\tau), \phi(t+\tau)] \rangle \langle n[\alpha, \beta, \theta(t), \phi(t)] \rangle} 
\label{autocorr}
\end{align}

In order to reproduce the experimental results shown in Figs.~\ref{Fig3} and~\ref{Fig4} of the main text, we assume the following normalized probability distribution for the initial polarization of the polariton laser:

\begin{align}
p_{init}(\theta, \phi)=\frac{\sin \theta}{4 \pi \frac{\sinh \Delta}{\Delta}} \times \exp(\Delta \cos(\theta)).
\label{pinit}
\end{align}

This distribution favors the formation of the polariton laser in the Vertical over Horizontal polarization for values of $\Delta>0$, and it assumes equal probability for D, A, L or R components of the initial polarization. The value of $\Delta$ is related to the ratio between Vertical and Horizontal polarization probabilities and, thus to the linear polarization splitting of the system. 

With this initial probability distribution we can calculate the initial average density along the detector axis given by $\alpha$ and $\beta$: 

\begin{align}
\langle n_{init}[\alpha, \beta] \rangle = \int_0^{\pi} d\theta \int_0^{\pi} d\phi p_{init}(\theta, \phi) n(\alpha, \beta, \theta, \phi)  
\label{initdist}
\end{align}

A similar calculation can be done to obtain the numerator in Eq.~(\ref{autocorr}), resulting in a zero-delay value of the measured autocorrelation function:

\begin{widetext}
	\begin{align}
	g^{(2)}(\tau=0,\alpha, \beta)=\frac{(\Delta^2+3)\cos(2\alpha)-4\Delta \cos(\alpha)+\Delta \coth(\Delta)[4\Delta \cos(\alpha)-3 \cos(2\alpha)-1]+3\Delta^2 +1}{2\Delta^2\left(-\frac{\cos(\alpha)}{\Delta} + \cos(\alpha)\coth(\Delta)+1\right)^2}
	\label{autocorrpolar}
	\end{align}
\end{widetext}

The auto-correlation function  at zero time delay [Eq.~(\ref{autocorrpolar})] depends only on $\alpha$ (choice of detected polarization) and $\Delta$. In the experiments shown in Fig.~2, we do not measure $g^{(2)}(\tau=0)$ at $t=0$ but at $t_{max}$, few tens of picoseconds after the beginning of lasing. We have checked that the displayed experimental results do not change significantly for earlier $t$. 
The measured values of $g^{(2)}(t_{max},\tau=0)$ for all polarization axis are well reproduced by a single fitting parameter $\Delta=1.4$, as shown in Fig.~\ref{Fig3}. This value of $\Delta$ results in a probability 4.5 times larger for the initial state to be V polarized (initial state contained in the upper hemisphere of the Poincar\'e sphere, Fig.~\ref{FigS1}(c)) than to be H polarized (initial state contained in the lower hemisphere of the Poincar\'e sphere).

\section{Estimation of the interaction energy}
\label{AppII}

We can evaluate the magnitude of the polariton interactions both in the planar cavity and the micropillar by estimating the number of polaritons in the lasing mode. To do so, we assume that exactly at threshold, each spatial mode is populated by one polariton. In the case of the planar cavity, at $P=1.5 P_{th}$ the measured intensity shows that we have 7 polaritons per mode (the emitted intensity is 7 times stronger than at threshold). We can estimate the size of each spatial modes from the measured polariton lifetime, given by the measured Q factor, and the polariton mass. In our case, the polariton lifetime is $30 \, \mathrm{ps}$, while its mass is $1\cdot10^{-4}m_e$, giving an estimated spatial diameter of 5.8~$\mu \mathrm{m}$ for the $k=0$ modes. Thus we find that the number of modes within the lasing region, with a diameter of $15\, \mu \mathrm{m}$ (determined by the excitation spot), is about 7.

Assuming a polariton-polariton interaction constant of $\alpha_{1}^{2D}=2 \,\mu \mathrm{eV} \cdot \mu \mathrm{m}^2$~ \cite{Ferrier2011} and accounting for a mode diameter of 5.8 $\mu \mathrm{m}$, we estimate the total interaction energy at $P=1.5 P_{th}$ to be $\alpha_{1}^{2D} (|\Psi_L|^2+|\Psi_R|^2)= 0.54 \,\mu \mathrm{eV}$ (where we have neglected interaction between polaritons with opposite circular polarization~\cite{Vladimirova2010,Paraiso2011, Takemura2014}).

A similar estimation of the interaction energy can be done for the pillar cavity. In this case the mode size is determined by the pillar size, and amounts to $7\,\mu \mathrm{m}^2$. By comparing the emitted intensity at $P_{th}$ and $P=5 P_{th}$, we find 305 polaritons per mode at the latter power, resulting in an estimated interaction energy of $\alpha_{1} (|\Psi_L|^2+|\Psi_R|^2)= 85 \,\mu \mathrm{eV}$.

\bibliographystyle{prb}

\end{document}